\documentclass[11pt]{article}
\makeatletter
\@addtoreset{equation}{section}
\makeatother

\usepackage{amssymb}
\usepackage{cite}

\newcommand{\half}{\frac12}

\newcommand{\angles}[3]%
{\left\langle
  #1
  \vphantom{#2 #3}
  \left|
    \vphantom{#1 #3}
    #2
  \right|
  #3
  \vphantom{#1 #2}
\right\rangle
}

\newcommand{\Tr}{\mathop{\mbox{Tr}}\nolimits}

\renewcommand{\arraystretch}{1.5}
\begin{document}
\begin{titlepage}
\title{Spatial Geometry of Non-Abelian Gauge Theory \\
  in \(2 + 1\) Dimensions}

\author{Michel Bauer\thanks{
    Service de Physique Theorique, Saclay, 91191 Gil-sur-Yvette
    cedex, France.} \and
  Daniel Z. Freedman\thanks{Department of Mathematics and Center
    for Theoretical Physics, Massachusetts Institute of
    Technology, Cambridge, MA 02139.} \thanks{Research supported
    by NSF Grant \# PHY--9206867.\protect\newline\strut\protect\newline
    MIT-CTP \# 2430, May 1995.}
  }
\date{}
\end{titlepage}

\maketitle

\begin{abstract}
  The Hamiltonian dynamics of \(2 + 1\) dimensional Yang-Mills
  theory with gauge group SU(2) is reformulated in gauge
  invariant, geometric variables, as in earlier work on the \(3 +
  1\) dimensional case.  Physical states in electric field
  representation have the product form \(\Psi_{\mathrm{phys}}
  [E^{a i}] = \exp ( i \Omega [ E ] / g ) F [G_{ij}]\), where the
  phase factor is a simple local functional required to satisfy
  the Gauss law constraint, and \(G_{ij}\) is a dynamical metric
  tensor which is bilinear in \(E^{a k}\).  The Hamiltonian
  acting on \(F [ G_{ij} ]\) is local, but the energy density
  is infinite for degenerate configurations where \(\det G (x)\)
  vanishes at points in space, so wave functionals must be
  specially constrained to avoid infinite total energy.  Study of
  this situation leads to the further factorization \(F [G_{ij} ]
  = F_c [ G_{ij} ] \mathcal R [ G_{ij} ]\), and the product
  \(\Psi_c [E] \equiv \exp (i \Omega [ E ] / g ) F_c [G_{ij}]\)
  is shown to be the wave functional of a topological field
  theory.  Further information from topological field theory may
  illuminate the question of the behavior of physical gauge
  theory wave functionals for degenerate fields.
\end{abstract}

\section{Introduction} \label{sec:intro}

The purpose of this paper is to apply a recently developed [1]
spatial-geometric approach to SU(2) and SU(3) Yang-Mills theory
in $3+1$ dimensions to the case of the SU(2) gauge theory in
$2+1$ dimensions.  The Hamiltonian for states satisfying the
Gauss law constraint is simpler in the $2+1$ dimensional case,
and we can begin to consider its physical implications.

The approach combines the following two ideas: 1) Hamiltonian
dynamics can be reformulated in gauge invariant variables, with
the Gauss law constraint automatically satisfied, and 2)~the
basic equations of the canonical formalism, except the definition
of $H$, are invariant under diffeomorphisms of the spatial
domain.

Using these ideas one can show that physical states in electric
field representation \cite{GJ1} take the form
\begin{equation}
  \label{eq:Psi}
  \Psi
  \left[
    E^{ai}
  \right] = e^{i \Omega [E] / g} F
  \left[
    G_{ij}
  \right]
\end{equation}
where $\Omega [E]$ is an explicit local functional of $E^{ai}(x)$
required by the constraint, and $G_{ij}(x)$ is the positive gauge
invariant variable
\begin{equation}
  \label{eq:Gsubij}
  G_{ij} = \epsilon_{ik} \epsilon_{jl} E^{ak} E^{al}
\end{equation}
which transforms as a covariant 2-tensor under diffeomorphisms.

We derive an expression for the expectation value (or matrix element)
of the Hamiltonian in which the phase factor of (\ref{eq:Psi}) cancels.  In
such matrix elements
\begin{equation}
  \label{eq:langle}
  \angles{F
    \left[ G_{ij}
    \right]}{H}{F
    \left[ G_{ij}
    \right]}
\end{equation}

\begin{enumerate}
  \renewcommand{\labelenumi}{
    \alph{enumi})}
\item All gauge indices completely contract out.

\item $H\/$ contains covariant spatial derivatives $\nabla_i\/$
  with the Christoffel connection $\Gamma^k_{ij}(G)$ of the
  dynamical metric $G_{ij}$, and the curvature scalar $R(G)$ also
  appears.  This means that the underlying spatial geometry is
  purely Riemannian. (This was also true [1] for gauge group
  SU(2) in $3+1$ dimensions, while for SU(3) in $3+1$ dimensions
  a more complicated geometry with torsion was found).

\item $H\/$ also contains the fixed Cartesian metric
  $\delta_{ij}\/$ of $\mathbb R^2$.  So $H\/$ is {\em not\/}
  diffeomorphism invariant, but there is a clear separation of
  invariant and non-invariant parts.

\item The resolution of the Gauss law constraint requires the
  $1/g\/$ factor in (\ref{eq:Psi}) when the usual ``perturbative
  normalization" of $A\/$ and $E\/$ is used, and $H$ also
  contains $1/g\/$ and $1/g^2\/$ terms, as well as positive powers
  up to $g^2$.  It is therefore hard to see how to apply
  perturbation theory and the strong coupling expansion is also
  problematic.
\end{enumerate}
What has been achieved, therefore, is a reduction of the gauge
theory to the subspace of physical states, where we find a local
non-linear Hamiltonian in the three components of $G_{ij}(x)$
rather than the initial six components of $E^{ai}(x)$. But this
result may well be of only formal significance if appropriate
dynamical methods cannot be found.

There is one aspect of the new formulation which indicates that
the geometric structure may have physical implications.
Specifically, the Hamiltonian is singular for degenerate
configurations in which \( \det G (x) \) vanishes in the spatial
domain.  At the points of degeneracy the rank of $3 \times 2$
electric field matrix \( E^{a i} (x)\) is less than two.
Generically degeneracy occurs at isolated points in $\mathbb
R^2$.  Since \( H\/\) is the transform of the standard form
\begin{math}
  \int d^2x\,
  \left( E^2 + B^2
  \right)
\end{math},
the singularities are repulsive, and a variational trial
functional or candidate Schr\"odinger eigenfunction which is not
specially behaved for degenerate fields will have infinite
energy.  This does not necessarily mean that wave functionals
vanish, because some singularities are due to choice of variables
and are resolved without physical consequence.  However in this
case the singularity originates in the phase \( \Omega [E]\/\)
which is required by the Gauss law.  Heuristic arguments indicate
that the singularities are significant, but do not prove that wave
functions vanish.  However, we are led to examine the situation
more carefully in the context of a physical picture based on an
analogy with the centrifugal barrier in quantum mechanics.

It is a familiar fact that eigenfunctions for angular momentum \(
\ell \neq 0\/\) in a central potential take the form
\begin{equation}
  \label{eq:Psimell}
  \Psi^m_\ell ( \vec x ) =  Y^m_\ell (\hat x ) r^\ell R (r)
\end{equation}
of a product of spherical harmonic, the centrifugal factor \(
r^\ell\/\) and a regular function.  Our considerations suggest
the analogous product form
\begin{equation}
  \label{eq:PsiE}
  \Psi [E] = e^{i \Omega [E] / g} F_c [G] {\cal R} [G]
\end{equation}
for all physical wave functionals in the non-abelian gauge theory
in which the centrifugal functional \( F_c [G]\/\) carries
effects of the singularities of $H$, and the residual factor \(
{\cal R}
[G]\/\) is presumably unconstrained.  It should be pointed out
that there is no analogue in the field theory of \( \ell = 0\/\)
states which are not suppressed at \( r = 0\).  All physical
states carry the phase factor in (\ref{eq:PsiE}).

Comparison of the Hamiltonian (\ref{eq:langle}) with its quantum
mechanical analogue motivates the definition of \( F_c [G]\/\) as
the solution of a set of functional differential equations which
are exactly the diffeomorphism and Wheeler~de Witt constraint
equations of (the Euclidean continuation of) \( 2 + 1\/\)
dimensional gravity.  It then turns out that the product of the
first two factors in (\ref{eq:PsiE}) satisfies the constraint
equations of the Chern-Simons (or topological \( b/F\/\)), form
\cite{AT1} of the gravity theory.  The solution of these
constraints can be written as the path integral representation
\begin{eqnarray}
  \Psi_c [E] & \equiv & e^{i \Omega [E] / g} F_c [G] \nonumber \\
  \label{eq:Psic}
  & = & \int [ dU (x) ] \exp
  \left\{ \frac 1g \int d^2 x\: {\rm Tr}\,
    \left( E^i U^{-1} \partial_i U
    \right)
  \right\}
\end{eqnarray}
in which \( U (x)\/\) is an SU(2) matrix and \( e^i (x) = T^a
E^{a i} (x)\).  The phase factor can be extracted from
(\ref{eq:Psic}) and another path integral representation
written for the real gauge invariant factor \( F_c [G]\); see
(\ref{eq:mani}) below.

The key question is whether \( \Psi_c [E]\/\) vanishes for
degenerate fields.  Despite considerable study of the
representation (\ref{eq:Psic}) we have not so far been able to
answer this question.  Another approach is to insert the product
representation (\ref{eq:PsiE}) in the Hamiltonian, and examine
the constraints on the residual factor \( {\cal R} [G]\/\) for
singular metrics.  The result is that the Hamiltonian governing
\( {\cal R} [G]\/\) is considerably less singular than the form
(\ref{eq:langle}), so that \( {\cal R} [G]\/\) need not vanish
for degenerate fields whether or not \( F_c [G]\/\) vanishes.

At present the conclusions are rather ambiguous.  Heuristic
analysis of the singularities of \( H\/\) suggests that physical
wave functions \( \Psi [E]\/\) vanish for degenerate fields, but
this is not confirmed by further investigation.  Instead there is
an apparently consistent scenario in which the factor \( \Psi_c
[E]\/\) of (\ref{eq:PsiE}) resolves the singularities of \( H\/\)
without the requirement that either \( \Psi_c [E]\/\) or \( {\cal
R} [G]\/\) vanish.  Thus the factor \( \Psi_c [E]\/\) plays an
important role in both scenarios, and it is very curious that the
wave functional of a topological theory enters into the analysis
of a non-trivial dynamical theory.  Of course, the explicit form
of \( \Psi_c [E]\/\) should settle the issue of its vanishing,
and it is to be hoped that there is now sufficient knowledge of
two dimensional and topological field theory to make progress on
this.

The previous discussion raises the question whether any similar
situation is expected in \( 3 + 1\/\) dimensions.  The
Hamiltonian of [1] is considerably more complicated, but it is
again singular for degenerate configurations of a tensor variable
bilinear in the electric field.  A careful study of the
significance of these singularities is required.  However even a
cursory inspection of the Hamiltonian shows that physical wave
functionals naturally have the product structure (\ref{eq:PsiE})
with a prefactor \( \Psi_c [E]\/\) which satisfies the constraint
equations of a 4-dimensional topological b-F theory \cite{H2},
and that the solution of these equations is just (\ref{eq:Psic})
again, but with integration over $\mathbb R^3\/$ instead of
$\mathbb R^2$.

The first gauge invariant formulation of Yang-Mills theory was
obtained by Halpern for the self dual theory, and a metric tensor
also appeared in his work \cite{H1}.  The gauge non-invariant
metric tensor $G_{\mu \nu} = A^a_\mu A^a_\nu\/$ was used as the
effective field variable in the long distance limit by Ne'eman
and Sijacki \cite{NS1}.  Lunev has developed a geometric
formulation of the Lagrangian form of non-abelian gauge theories
both for $2 + 1\/$ and $3 + 1\/$ dimensions \cite{L1}.
Discussions of a gauge field geometry with torsion have
also appeared recently \cite{FHJL1}, \cite{HMO1}, and there is a
geometric formulation of the Hamiltonian dynamics of $3 + 1\/$
dimensional SU(2) gauge theory which uses the potential
representation \cite{HJ1}.

Variational calculations for gauge theory in which the Gauss law
constraint is enforced by averaging the gauge group have been
presented by Kogan and Kovner \cite{KK1}.  We also list here some
other references to recent studies of the non-perturbative
physics of non-abelian gauge theories in \(2 + 1\) and \(3 + 1\)
dimensions \cite{AFF1,L2,LMT1,CLG1,DW1,MV1}.

\section{Spatial Geometry of the Gauge Theory}
\label{sec:2}

\subsection{The Canonical Formalism in $E\/$-field
  Representation}
\label{subsec:can}

The action\footnote{The perturbative ``normalization'' of gauge
fields was implicitly used in Sec.~\ref{sec:intro}, but we now
use the normalization in which \( 1 / g^2\/\) appears as a factor
in the action.  The scaling \( A \to g A\/\) and \( E \to E /
g\/\) may be used to change any formula in
Sections~\ref{subsec:can}--\ref{sec:6} to perturbative
normalization.}
of SU(2) gauge theory in \( 2 + 1\/\) dimensional flat
Minkowski space is
\begin{equation}
  \label{eq:14g2}
  S = - \frac{1}{4 g^2} \int d^2x\, dt (F_{\mu \nu}^a)^2.
\end{equation}
The coupling constant has dimensions of inverse length, \( [g] =
1\).  This gives a theory which is super-renormalizable in
perturbation theory, but there are still unresolved
non-perturbative issues \cite{F1}, namely the questions of
confinement and generation of a mass gap.

We wish to set up the canonical formalism in \( A^a_0 = 0\/\)
gauge.  The canonical momentum is \( E^{ai} (x) = \delta S /
\delta \dot{A}^a_i (x) = - F^{a (0 i)} / g^2\), and the canonical
commutation rule, Gauss law constraint, magnetic field and
Hamiltonian are
\begin{eqnarray}
  \label{eq:2.2}
  \left[
    {A^a}_j (x), E^{b k}
    \left(
      x'
    \right)
  \right] & = & i \delta^{a b} {\delta_j}^k \delta^{(2)} (x -
  x') \\
  \label{eq:2.3}
  G^a (x) \psi & = &
  \left(
    \partial_i E^{a i} (x) + \epsilon^{a b c} {A^b}_i (x) E^{c i}
    (x)
  \right) \psi = 0 \\
  \label{eq:2.4}
  B^a (x) & = & \epsilon^{i j}
  \left(
    \partial_i {A^a}_j + \half \epsilon^{a b c} {A^b}_i
    {A^c}_j
  \right) \\
  \label{eq:2.5}
  H & = & \half \int d^2 x
  \left\{
    g^2 \delta_{i j} E^{a i} E^{a j} + \frac{1}{g^2} B^a B^a
  \right\}
\end{eqnarray}
Only the definition of \( H\/\) requires the Cartesian spatial
metric \( \delta_{i j}\), and (\ref{eq:2.2}--\ref{eq:2.4}) are
covariant under diffeomorphisms of the spatial domain \(
\mathbb R^2\), that is coordinate transformations \( x^i \to
y^\alpha (x^i)\), \( i, \alpha = 1, 2\/\) and transformation
rules
\begin{equation}
  \label{eq:2.6}
  \everymath{\displaystyle}
  \begin{array}{rcl}
    {A^a}_i (x) & \to & {A^a}_\alpha (y) =
    \frac{\partial x^i}{\partial y^\alpha} {A^a}_i (x) \\
    E^{a i} (x) & \to & E^{a \alpha} (y) =
    \left|
      \frac{\partial x}{\partial y}
    \right| \frac{\partial y^\alpha}{\partial x^i} E^{a i} (x)
  \end{array}
\end{equation}
from which we see that \( A\/\) is a covariant vector (a 1-form)
and \( E\/\) is a contravariant vector density.

The dynamical problem of gauge theory is to find solutions of the
functional Schr\"odinger equation \( H \psi = E \psi\/\) for
states which satisfy (\ref{eq:2.3}).  We note that \( H\/\) is
not diffeomorphism invariant, because the fixed metric \(
\delta_{ij}\/\) appears and because both terms have density
weight two in the dynamical variables, whereas weight one is
required for invariance.  Nevertheless we shall be guided in our
work by the idea of preserving the diffeomorphism covariance of
the canonical formalism.

In almost all work on the Hamiltonian formalism in gauge theory,
the potential representation is used in which \( \psi \to \psi
[A]\/\) and the electric field acts by functional differentiation
\( E^{a j} = - i \delta / \delta {A^a}_j\).  However the
implementation of Gauss' law (\ref{eq:2.3}) leads either to a
non-local Hamiltonian \cite{HJ1,LMT1} or to averaging over the
gauge group using additional variables \cite{KK1}.  We therefore
use the electric field representation \cite{GJ1} obtained by the
canonical Fourier transformation
\begin{equation}
  \label{eq:2.7}
  \psi [A] =
  \int
  \left[
    d E^{a i} (x)
  \right] \exp
  \left\{
    i \int d^2 x {A^a}_i (x) E^{a i} (x)
  \right\} \Psi [E].
\end{equation}
In this representation it is \( {A^a}_j = i \delta / \delta E^{a j}\/\)
which acts by differentiation.

We now consider the gauge transformation by the \( 3 \times 3\/\)
orthogonal matrix \( T^{a b} (x)\), which acts as
\begin{equation}
  \label{eq:2.8}
  \begin{array}{rcl}
    {A^a}_i & \to & T {A^a}_i = \half \epsilon^{a b c} T^{b d}
    \partial_i T^{c d} + T^{a b} {A^b}_i \\
    E^{a i} & \to & T E^{a i} = T^{a b} E^{b i}
  \end{array}
\end{equation}
The Gauss law (\ref{eq:2.3}) requires \( \psi [T A] = \psi
[A]\/\) and this gives \cite{GJ1}
\begin{equation}
  \label{eq:2.9}
  \psi [T E] = \exp -i \int d^2 x \half \epsilon^{a b c} E^{a i}
  \left(
    T ^{d b}\partial_i T^{d c}
  \right)\: \psi [E].
\end{equation}
It is the fact that the convective term in the Gauss law is a
multiplication operator in \( E\/\)-field representation, viz
\begin{equation}
  \label{eq:2.10}
  G^a (x) \psi [E] = \frac 1 g
  \left(
    \partial_i E^{a i} (x) - i \epsilon^{a b c} E^{b c} (x)
    \frac{\delta}{\delta E^{c i} (x)}
  \right) \psi [E]
\end{equation}
that leads to the phase factor in (\ref{eq:2.9}).

\subsection{The Unitary Transformation}
\label{subsec:unit}

In the same spirit as in \cite{GJ1}, but with some differences,
we implement (\ref{eq:2.9}) by a unitary transformation
\begin{equation}
  \label{eq:2.11}
  \psi [E] = \exp i \Omega [E]\: F [E]
\end{equation}
in which the phase factor \( \exp i \Omega [E]\/\) is the
intertwining operator which removes the convective term from the
Gauss law generator:
\begin{equation}
  \label{eq:2.12}
  \everymath{\displaystyle}
  \begin{array}{rcl}
    G^a (x) \exp i \Omega [E] & = &
    \exp i \Omega [E]\: \overline{G}^a (x) \\
    \overline{G}^a (x) & \equiv &
    - i \epsilon^{a b c} E^{b j} (x)
    \frac{\delta}{\delta E^{c j} (x)}
  \end{array}
\end{equation}
It is clear that (\ref{eq:2.9}) is satisfied by any functional \(
\Omega [E]\/\) with the gauge transformation property
\begin{equation}
  \label{eq:2.13}
  \Omega [T E] = \Omega [E] - \half \int d^2 x E^{a i}
  \epsilon^{a b c} T^{d b} \partial_i T^{d c}
\end{equation}
and we can see that (\ref{eq:2.12}) is also satisfied by looking
at the form of (\ref{eq:2.13}) for infinitesimal gauge
transformations.  We also require that \( \Omega [E]\/\) be
invariant under spatial diffeomorphisms.

The phase \( \Omega [E]\/\) which is introduced to satisfy Gauss'
law is also the key to the spatial geometric properties of our
approach to gauge theory.  Most of these properties can be
deduced directly from the gauge (\ref{eq:2.13}) and
diffeomorphism requirements for \( \Omega [E]\), rather than from
any specific form.  Nevertheless it is useful to exhibit the
following simple local functional which is easily shown to
satisfy both requirements:
\begin{equation}
  \label{eq:2.14}
  \Omega [E] = \int d^2 x \epsilon^{a b c}
  \left(
    E^{a i} E^{b j} \partial_i E^{c k}
  \right) \varphi_{j k}
\end{equation}
where \( \varphi^{i j} = E^{a i} E^{a j}\/\) is a gauge invariant
tensor density and \( \varphi_{j k}\/\) is its matrix inverse
(i.e., \( \varphi^{i j} \varphi_{j k} = {\delta^i}_k\)).  One
should note that the two requirements do not specify \( \Omega
[E]\/\) uniquely, but that any two solutions (e.g., \( \Omega
[E]\/\) and \( \Omega' [E]\/\)) must differ by a gauge and
diffeomorphism invariant functional.  For example, one could
take
\begin{equation}
  \label{eq:2.15}
  \Omega' [E] = \Omega [E] + c \int d^2 x
  \left(
    \det \varphi^{m n}
  \right)^{1/2}
\end{equation}
and there are many other possibilities.  It turns out that the
choice (\ref{eq:2.14}) gives the simplest spatial geometry in a
way which will make precise below, but we now resume the general
discussion which is independent of any specific choice.

The next steps in the development of the geometry are

\noindent 1) implementation of the unitary transformation on the
operators of the theory, specifically
\begin{equation}
  \label{eq:2.16}
  \everymath{\displaystyle}
  \begin{array}{rcl}
    \overline E^{a i} (x) & = &
    e^{-i \Omega [E]} E^{a i} (x) e^{i \Omega [E]} \\
    {\overline A^a}_i (x) & = &
    e^{- i \Omega [E]} {A^a}_i (x) e^{i \Omega [E]},
  \end{array}
\end{equation}
2) study of the residual Gauss constraint on \( F [E]\),
\begin{equation}
  \label{eq:2.17}
  \overline G^a (x) F [E] = 0
\end{equation}
which implies \(F [TE] = F [E]\) for finite gauge
transformations, and

\noindent 3) expression of the Hamiltonian in geometric
variables.

The transformed operators are
\begin{equation}
  \label{eq:2.18}
  \everymath{\displaystyle}
  \begin{array}{rcl}
    \overline E^{a j} (x) & = & E^{a j} (x) \\
    {\overline A^a}_j (x) & = & i
    \frac{\delta}{\delta E^{a j} (x)} -
    \frac{\delta \Omega}{\delta E^{a j} (x)} \\
    & \equiv & i \frac{\delta}{\delta E^{a j} (x)} +
    {\omega^a}_j (x).
  \end{array}
\end{equation}
The quantity \( {\omega^a}_i (x)\/\) is a covariant vector if \(
\Omega [E]\/\) is diffeomorphism invariant, while (\ref{eq:2.13})
tells us that
\begin{equation}
  \label{eq:2.19}
  {\omega^a}_i \to \half \epsilon^{a b c} T^{b d} \partial_i
  T^{c d} + T^{a b} {\omega^b}_i
\end{equation}
under the gauge transformation \( E^{a i} \to T^{a b} E^{b i}\).
These are exactly the properties of an SU(2) gauge connection, so
the unitary transformation produces the composite gauge
connection \( {\omega^a}_i (x)\/\) which depends on the electric
field.  If the specific phase \( \Omega [E]\/\) of
(\ref{eq:2.14}) is used, we see that \( {\omega^a}_i (x)\/\) is a
local function of \( E\/\) and its first derivatives.

The transformed magnetic field is
\begin{eqnarray}
  \overline B^a & = &
  \epsilon^{j k}
  \left(
    \partial_j \overline A_k + \half \epsilon^{a b c}
    {\overline A^b}_j {\overline A^c}_k
  \right) \nonumber \\
  \label{eq:2.20}
  & = & \widehat B^a + i \epsilon^{j k} \widehat D_j
  \frac{\delta}{\delta E^{a k}} - \half \epsilon^{j k}
  \epsilon^{a b c} \frac{\delta^2}{\delta E^{b j} \delta E^{c k}}
\end{eqnarray}
where (\ref{eq:2.18}) has been used to get the second line which
contains the gauge covariant derivative with connection \(
\omega\/\) and the composite magnetic field
\begin{eqnarray}
  \label{eq:2.21}
  \hat D_i \frac{\delta}{\delta E^{a j}} & \equiv &
  \partial_i \frac{\delta}{\delta E^{a j}} + \epsilon^{a b c}
  {\omega^b}_i \frac{\delta}{\delta {E^c}_j} \\
  \label{eq:2.22}
  \widehat B^a & \equiv &
  \epsilon^{i j}
  \left(
    \partial_i {\omega^a}_j + \half \epsilon^{a b c} {\omega^b}_i
    {\omega^c}_j
  \right).
\end{eqnarray}
We have dropped the contact term
\begin{equation}
  \label{eq:2.23}
  \epsilon^{i j} \epsilon^{a b c}
  \frac{\delta}{\delta E^{b i} (x)} {\omega^c}_j (x)
\end{equation}
in (\ref{eq:2.20}), which contains the ill-defined quantities \(
\delta (0)\/\) and \( \partial_i \delta (0)\/\) if
(\ref{eq:2.14}) is used.  The covariant point splitting argument
of \cite{MB1} shows that this contact term actually vanishes.  In
general, however, the important issue of the regularization
within our approach to gauge theory has not yet been studied.

With \( \overline B^{a i}\/\) as given in (\ref{eq:2.20}) the
unitary transform of the Hamiltonian (\ref{eq:2.5}) may be
written as
\begin{equation}
  \label{eq:2.24}
  \overline H = \half \int d^2 x
  \left\{
    g^2 \delta_{ij} E^{a i} E^{a j} + \frac{1}{g^2}
    \overline B^a \overline B^a
  \right\}.
\end{equation}
This may be regarded as an intermediate result within our
approach.  It is gauge covariant, because of the placement of the
composite connection \( {\omega^a}_i\), but spatial geometric
variables have not yet appeared.

Finally we note two useful identities which follow from the
definition (\ref{eq:2.18}) of the composite connection.  For an
infinitesimal gauge transformation \( T^{a b} (x) = \delta^{a b}
- \epsilon^{a b c} \theta^c (x)\), the gauge requirement
(\ref{eq:2.13}) on the phase \( \Omega [E]\/\) reduces to
\begin{equation}
  \label{eq:2.25}
  \int d^2 x E^{a i}
  \left(
    \partial_i \theta^a + \epsilon^{a b c} {\omega^b}_i \theta^c
  \right) = 0.
\end{equation}
Partial integration gives the gauge identity
\begin{equation}
  \label{eq:2.26}
  \widehat D_i E^{a i} \equiv \partial_i E^{a i} +
  \epsilon^{a b c} {\omega^b}_i E^{c i} = 0.
\end{equation}
We also consider the effect of an infinitesimal diffeomorphism of
\( E\/\) parameterized by the vector field \( v^i (x)\).  The
transformation law becomes
\begin{equation}
  \label{eq:2.27}
  \delta E^{a i} =
  \left(
    \partial_j v^i
  \right) E^{a j} - \partial_j
  \left(
    v^j E^{a i}
  \right).
\end{equation}
But \( \Omega [E]\/\) is required to be invariant, so that
\begin{equation}
  \label{eq:2.28}
  0 =
  \int d^2 x \frac{\delta \Omega}{\delta E^{a i}} \delta E^{a i}
  = - \int d^2 x v^j
  \left[
    E^{a i}
    \left(
      \partial_i \omega_j - \partial_\nu \omega_i
    \right) + {\omega^a}_j \partial_i E^{a c}
  \right].
\end{equation}
Inserting the gauge identity (\ref{eq:2.25}), and using
(\ref{eq:2.22}) one finds the diffeomorphism identity
\begin{equation}
  \label{eq:2.29}
  \epsilon_{i j} E^{a j} \widehat B^a = 0
\end{equation}
which ensures orthogonality the electric field and the composite
magnetic field.

\subsection{The functional $F[E]$}

Our goal in this section is to solve explicitly the Gauss law
constraint for the functional $F[E]$, which according to
(\ref{eq:2.8},\ref{eq:2.17}) is simply $F[TE] = F[E]$.  If
$E=\{E^{ai}(x)\}$ is a configuration of the electric field, the
gauge transformed configuration is $TE = \left\{ T^{ab}(x)
E^{bi}(x) \right\}$. This transformation is purely local, and we
shall analyze it point by point, without mentioning $x\/$
explicitly. Then the tensor index of the electric field is just a
label.  We view $E^{a1}\/$ and $E^{a2}\/$ as two vectors in
(gauge) 3-space. Classical invariant theory then asserts that the
invariants are freely generated by the scalar products
$\varphi^{ij} \equiv E^{ai} E^{aj}$. There is a quick proof if
one assumes that $E^{a1}$ and $E^{a2}\/$ are linearly
independent, or equivalently that $\det \varphi^{ij} \neq 0$. We
apply Schmidt orthogonalization to write $E^{a1} =
\lambda_1n^{a1}\/$ with $\lambda_1 > 0$ and $n^{a1}\/$ a unit
vector, and then $E^{a2} = \mu_2n^{a1} + \lambda_2n^{a2}\/$ with
$\lambda_2 > 0$ and $n^{a2}\/$ a unit vector orthogonal to
$n^{a1}$. Then $(n^{a1}, n^{a2}, n^{a1} \times n^{a2})$ is an
orthonormal frame. The numbers $\lambda_1$, $\lambda_2$ and
$\mu_2$ are uniquely defined in terms of $\varphi^{ij}\/$ and the
correspondence is one to one. There is a unique rotation sending
$n^{a1}\/$ to $(1,0,0)$ and $n^{a2}\/$ to $(0,1,0)$. This
rotation sends $E^{a1}\/$ to $(\lambda_1,0,0)$ and $E^{a2}\/$ to
$(\mu_2, \lambda_2, 0)$. So we have found a canonical
representative depending only on the invariants $\varphi^{ij}\/$
on each non-degenerate orbit. This is the result we were after.
Wilson lines for the composite connection ${\omega^a}_i$, which are
gauge invariant objects, can be expressed solely in terms of
$\varphi^{ij}$. The point is that the Wilson line is unchanged
when one replaces a given electric field configuration by the
canonical representative (depending only on $\varphi^{ij}$) of
its gauge orbit.

To summarize we have shown that the most general gauge invariant
functional of the electric field can be rewritten as a
functional of $\varphi^{ij}$ or equivalently of $G_{ij}$ defined
by (\ref{eq:Gsubij}) which as we shall see later is a more
convenient variable.  Hence on the physical subspace we can write
$F[G_{ij}]$ for the wave functional.

In two dimensions and planar or spherical geometry, the gauge
group is connected and to check for gauge invariance it is enough
to check infinitesimal gauge invariance (i.e., the Gauss law).

One can use the fact that the general \(E^{a i}\) field is the
gauge transform of its canonical representative to relate the
measures.
\begin{equation}
  \label{eq:2.29a}
  \prod_{a,i} d E^{ai} \propto d \rho_{\rm Haar}^{\rm SO(3)}
  \prod_{i \leq j} d G_{ij}.
\end{equation}

\subsection{Tensor variables}

We now wish to introduce a basis \({e^a}_1\), \({e^a}_2\),
\(e^a\) in the adjoint representation of SU(2).  These are
defined in terms of the electric field by
\begin{equation}
  \label{eq:2.29b}
  E^{ai} = \epsilon^{ij} e^a_j, \quad
  \epsilon^{abc} e^a_i e^b_j = \sqrt{G} \epsilon_{ij} e^c.
\end{equation}
It is easy to check that $e^a_i\/$ is a gauge vector and a
covariant vector, that the metric $G_{ij}$ of (\ref{eq:Gsubij})
is simply $e^a_i e^a_j$, a gauge invariant symmetric positive
covariant tensor whose determinant we denote by $G$, and that
$e^a\/$ is a (unit) gauge vector and a pseudoscalar. From now on,
we use the metric $G_{ij}$ to raise and lower indices, the only
exceptions being the epsilon symbols. The three gauge vectors
$e^a_1$, $e^a_2\/$ and $e^a\/$ are linearly independent and
satisfy
\begin{equation}
  \label{eq:2.29c}
  e^a_i e^a = 0,\quad
  e^{ai} e^a_j = \delta^i_j,\quad
  \delta^{ab} = e^a e^b + e^{ai} e^b_i.
\end{equation}
Any quantity with gauges indices can be expanded in the
semi-orthogonal frame ${e^a}_1$, ${e^a}_2$, $e^a$. For gauge
covariant quantities, the coefficients will be geometric objects
in the spatial geometry of the gauge theory.

\subsection{Riemannian geometry}

We expand the gauge covariant derivative of the basis
vectors. We can write
\begin{equation}
  \label{eq:2.29d}
  \hat{D}_i e^a_j \equiv
  \partial_i {e^a}_j + \epsilon^{a b c} {\omega^b}_i {e^c}_j
  =  M^l_{ij} e^a_l + N_{ij} e^a
\end{equation}
By computing the effect of a change of coordinates, one sees
explicitly that $N_{ij}$ is a tensor and $M^l_{ij}$ an affine
connection. The connection is metric compatible because
\begin{equation}
  \label{eq:2.29e}
  \partial_i G_{jk} = \hat{D}_i G_{jk} = \hat{D}_i (e^a_je^a_k) =
  M^l_{ij} e^a_le^a_k + M^l_{ik} e^a_l e^a_j = M^l_{ij} G_{kl} +
  M^l_{ik} G_{jl}.
\end{equation}

 From the gauge identity $\hat{D}_i E^{ai}=\epsilon^{ij}\hat{D}_i
e^a_j=0$ we conclude that $\hat{D}_i e^a_j$ is symmetric in $ij$,
and the same is true of $M^k_{ij}$ and $N_{ij}$. But the only metric
compatible symmetric affine connection is the Levi-Civita connection.
Hence
\begin{equation}
  \label{eq:2.29f}
  M^k_{ij} = \Gamma^k_{ij} \equiv \half G^{kl}
  \left(
    \partial_i G_{lj} + \partial_j G_{il} - \partial_l G_{ij}
  \right).
\end{equation}

 From $\epsilon^{ij}\hat{D}_i e^a_j=0$ we also get, by contraction with
$e^a_k\/$, another form of the gauge identity:
\begin{equation}
  \label{eq:2.29g}
  e^a_k\epsilon^{ij}\partial_i e^a_j -\sqrt{G}e^a \omega^a_k = 0.
\end{equation}

Using the Leibnitz property of the gauge covariant derivative we
compute $\hat{D}_i e^a=-{N_i}^k e^a_k $. To summarize, we have
proved that
\begin{equation} \label{eq : cov}
  \begin{array}{rcl}
    \hat{D}_i e^a_j & = & \Gamma^k_{ij} e^a_k + N_{ij} e^a \\
    \hat{D}_i e^a & = & -{N_i}^k e^a_k .
  \end{array}
\end{equation}
were $\Gamma^k_{ij}$ is the Levi-Civita connection and $N_{ij}$ a
symmetric tensor.

Up to now, we have only used general properties of the phase
$\Omega[E]$ and the outcome is already quite remarkable. But to
go further we need an explicit formula. We have chosen the
simplest possibility, for which $N_{ij}\/$ vanishes.

\subsection{The explicit phase $\Omega[E]$ and its variation}

The formula
\begin{equation}
  \label{eq:2.30a}
  \Omega[E]  = \int d^2\!x \sqrt{G}e^{ai}\partial_i e^a
\end{equation}
defines a scalar. It is not difficult to check that this
definition coincides with (\ref{eq:2.14}) and has the proper
behavior under gauge transformations (\ref{eq:2.13}).

It is clear that $\Omega[E]$ is a homogeneous function of the
electric field of degree 1, but it also has a deeper symmetry,
which we call local tensorial homogeneity. Let $\Lambda^i_j\/$ be
an invertible tensor field of type $(1,1)$.  We can use it to
define a local $GL(2)$ transformation on the electric field by
$(\Lambda E)^{ai} = \Lambda^i_j E^{aj}$. Using the original
formula (\ref{eq:2.14}) and inserting the frame variables after
the transformation, it is easy to check that
\begin{equation}
  \label{eq:2.30b}
  \Omega[\Lambda E] = -\int d^2\!x \sqrt{G} \Lambda^i_j
   e^a \partial_ie^{aj}.
\end{equation}
The infinitesimal version gives, after some reshuffling of indices,
\begin{equation}
  \label{eq:2.30c}
  \sqrt{G} \omega^a_i e^{ak} = e^a \epsilon^{jk} \partial_i
  e^a_j,
\end{equation}
an identity which bears a striking resemblance to the gauge
identity (\ref{eq:2.29g}).  Those two identities fix the scalar
product in gauge space of the composite connection with the three
basis vectors. One can see explicitly that
\begin{equation}
  \label{eq:2.30e}
  \sqrt{G} \omega^a_i = \epsilon^{jk} (e^a_k e^b \partial_i e^b_j
  + e^a e^b_i \partial_j e^b_k)
\end{equation}
and that
\begin{equation} \label{eq : cov1}
  \begin{array}{lcl}
    \hat{D}_i e^a_j & = & \Gamma^k_{ij} e^a_k \\
    \hat{D}_i e^a & = & 0 \\
    \hat{B}^a & = &-\half \sqrt{G} R e^a.
  \end{array}
\end{equation}

The first two equations show the vanishing of $N_{ij}$ for our choice of
phase. The third one comes from the computation of the commutator of the
covariant derivative. One finds
\begin{equation}
  \label{eq:2.31a}
  \left[
    \hat{D}_i, \hat{D}_j
  \right] e^a_k =
  R^l_{k \,ij}e^a_l
\end{equation}
But $\left[\hat{D}_i, \hat{D}_j \right] \lambda^a =
\epsilon^{abc} F^b_{ij} \lambda^c = \epsilon^{abc} \epsilon_{ij}
\hat{B}^b \lambda^c$. According to the diffeomorphism identity,
the composite magnetic field is perpendicular to the electric
field. Hence $\hat{B}^b = \hat{B}e^b\/$ and using various two
dimensional tensor identities, such as
\begin{equation}
  \label{eq:2.31b}
  2 R^m_{k\, ij} = -\det G \epsilon_{ij} \epsilon_{kl} G^{lm} R,
\end{equation}
where $R\/$ is the scalar curvature, and we finally
obtain
\begin{equation}
  \label{eq:2.31c}
  \hat{B} = -\half \sqrt{G} R.
\end{equation}

Some insight into the meaning of the connection~(\ref{eq:2.30e})
may be gained by examining it in the special gauge where
\({e^3}_i = 0\) and \(e^a = (0, 0, 1)\).  The remaining
components \({e^a}_i\) may be viewed as a standard frame
(zweibein) of a Riemannian 2-manifold, and one finds that
\({\omega^a}_i\) is related to conventional Riemannian spin
connection by \({\omega^a}_i = - \frac 12 \epsilon^{a b c}
{\omega^{b c}}_i\).

\subsection{The Hamiltonian}

Our next task is to calculate the expectation value of the
Hamiltonian~(\ref{eq:2.24}) in gauge invariant physical states \(F
[G_{i j}]\).  The non-trivial part is the action of the magnetic
field operator~(\ref{eq:2.20}), and we obtain this using the
geometric formulae~(\ref{eq : cov1}) together with the functional
chain rule
\begin{equation}
  \label{eq:2.47}
  \frac{\delta G_{mn}(x)}{\delta E^{bj}(y)}  = \epsilon_{mp}
  \epsilon_{nq}
  \left(
    \delta^p_j E^{bq}(x) + \delta^q_j E^{bp}(x)
  \right) \delta (x - y)
\end{equation}
which follows from~(\ref{eq:Gsubij}).

We start with the second derivative term in~(\ref{eq:2.20}),
which simplifies because \(\delta^2 G_{m n} / \delta E^{a i}
\delta E^{b j}\) is proportional to \(\delta^{b c}\) and does not
contribute.  Hence we get
\begin{equation}
  \label{eq:2.48}
  \epsilon^{abc}
  \frac{\delta^2 F}{\delta E^{bj}\delta E^{ck}} =
  \epsilon^{abc} \frac{\delta G_{mn}}{\delta E^{bj}}
  \frac{\delta G_{pq}}{\delta E^{ck}}
  \frac{\delta^2 F}{\delta G_{mn}\delta G_{pq}}.
\end{equation}
Elementary manipulations using~(\ref{eq:2.47}) and the
definition~(\ref{eq:2.29b}) of \(e^a\) then give
\begin{displaymath}
  -\half \epsilon^{i j} \epsilon^{a b c}
  \frac{\delta^2 F}{\delta E^{b i} \delta E^{c j}} =
  - 2 \sqrt G\, e^a \epsilon_{m p} \epsilon_{n q}
  \frac{\delta^2 F}{\delta G_{m n} \delta G_{p q}}.
\end{displaymath}
Note the determinant-like combination of second derivatives with
respect to \(G_{m n}\).

Turning to the gauge covariant derivative term
in~(\ref{eq:2.20}), we obtain using~(\ref{eq:2.29b}) and (\ref{eq:2.47})
\begin{equation}
  \label{eq:2.49}
  \epsilon^{i j} \hat D_i \frac{\delta F}{\delta E^{a j}} =
  - 2 \hat D_i
  \left(
    {e^a}_j \frac{\delta F}{\delta G_{i j}}
  \right).
\end{equation}
Using~(\ref{eq : cov1}) we then get
\begin{equation}
  \label{eq:2.50}
  \everymath{\displaystyle}
  \begin{array}{rcl}
  \epsilon^{i j} \hat D_i \frac{\delta F}{\delta E^{a j}} & = &
  -2 {e^a}_j
  \left(
    \partial_i \frac{\delta F}{\delta G_{i j}} + \Gamma^j_{i k}
    \frac{\delta F}{\delta G_{i j}}
  \right) \\
  & \equiv & -2 {e^a}_j \nabla_i \frac{\delta F}{\delta G_{i j}}.
  \end{array}
\end{equation}
It is curious but natural that the use of~(\ref{eq : cov1}),
which converts gauge geometry to spatial geometry, automatically
brings in the connection terms required for the diffeomorphism
covariant divergence of \(\frac{\delta F}{\delta G_{i j}}\).
This quantity formally transforms as a tensor density of weight
one.

Using~(\ref{eq : cov1}) again for \(\widehat B^a\), we arrive at
the geometrical formula for \(\overline B^a F\):
\begin{equation}
  \label{eq:2.51}
  \overline B^a F = - 2 i {e^a}_m \nabla_n
  \frac{\delta F}{\delta G_{m n}} - \sqrt G\, e^a
  \left(
    \half R F + 2 \epsilon_{m p} \epsilon_{n q}
    \frac{\delta ^2 F}{\delta G_{m n} \delta G_{p q}}.
  \right)
\end{equation}
A striking feature of this formula is that its real and imaginary
parts are orthogonal in gauge space.  A direct consequence is
that the energy density is real, as we will see below.  This
property was not manifest in the \(3 + 1\) dimensional cases treated
previously \cite{MB1}, and it should facilitate variational
calculations or lattice simulations.

It is now easy to write an explicit expression for expectation
values of \(\overline H\) in~(\ref{eq:2.24}):
\begin{equation}
  \label{eq:2.52}
  \angles F {\overline H} F =
  \int \left[ d G_{i j} \right] \int d^2 x\: (F [G] | \mathcal H
  | F [G])
\end{equation}
with functional measure~(\ref{eq:2.29a}) and energy density
\begin{equation}
  \label{eq:2.53}
  \everymath{\displaystyle}
  \begin{array}{rcl}
    (F [G] | \mathcal H | F [G]) & = &
    \half g^2 \delta^{i j} G_{i j} |F|^2 \\
    & & \quad \strut + \frac{2}{g^2} G_{i j} \nabla_k
    \frac{\delta F^*}{\delta G_{i j}} \nabla_\ell
    \frac{\delta F}{\delta G_{j \ell}} \\
    & & \quad \strut + \frac{\det G}{8 g^2}
    \left|
      R F + 4 \epsilon_{m p} \epsilon_{n q}
      \frac{\delta^2 F}{\delta G_{m n} \delta G_{p q}}
    \right|^2.
  \end{array}
\end{equation}

This spatial geometric form of the gauge theory Hamiltonian is
our principal result.  It is manifestly gauge invariant, real,
and local.  It is the sum of three positive definite
contributions, and the magnetic energy density is singular for
configurations where the metric degenerates.  The origin of these
singularities is the unitary transformation required by the Gauss
law constraint, and this is a non-perturbative effect.

\section{The singularities of \protect\boldmath$H$}
\label{sec:singularity}

The Hamiltonian derived in the previous section is the sum of
three real positive terms.  The last two terms are the
contribution of the magnetic energy density.  These terms involve
the Christoffel symbol (\ref{eq:2.29f}) and curvature scalar
\(R\), which are singular for space-dependent configurations of
\begin{math}
  G_{ij} (x)\/
\end{math}
which are degenerate, i.e.\
\begin{math}
  \det G (x) = 0
\end{math}.
Note that constant degenerate metrics do not make $H\/$ singular
because $G^{ij}\/$ is always multiplied by
\begin{math}
  \partial_j G_{k \ell}
\end{math}
in (\ref{eq:2.29f}).  In terms of the electric field
\begin{math}
  E^{a i} (x)
\end{math},
a degenerate configuration is entirely regular from a physical
standpoint.  What happens is that the vectors
\begin{math}
  E^{a1} (x)
\end{math}
and
\begin{math}
  E^{a2} (x)
\end{math}
become linearly dependent somewhere in space.  This is a gauge
invariant criterion.

Since the variable
\begin{math}
  G_{i  j} (x)\/
\end{math}
is a non-negative 2-tensor, any zero of $\det G (x)$ is
generically a local minimum.  This fact indicates that the
generic case of degeneracy occurs at isolated points of the
domain $\mathbb R^2$.  The same conclusion comes from the linear
dependence of the
\begin{math}
  E^{ai} (x)
\end{math}.
Given
\begin{math}
  E^{a1} (x)
\end{math}
and
\begin{math}
  E^{a 2} (x)
\end{math},
the conditions
\begin{math}
  E^{a1} (x) = c E^{a 2} (x)
\end{math}
constitute three equations to determine the three quantities
$x^1$, $x^2$, and $c$.  So again one expects that solutions occur
at isolated points.

Let us now exemplify the statement in the introduction that a
wave functional which is not specially constrained for degenerate
fields has infinite energy.  Consider the smooth non-covariant
functional
\begin{equation}
  \label{eq:FGwedge}
  F
  \left[ G_{i  j}
  \right] = \exp \int d^2 x
  \left\{
    - \half G_{11} \nabla^2 G_{11} - \half G_{22} \nabla^2 G_{22}
    - \delta^{i j} G_{i j}.
  \right\}
\end{equation}
This is normalizable, since
\begin{math}
  G^2_{12} \leq G_{11} G_{22}\/
\end{math},
is damped at short wavelengths by the flat Laplacian $\nabla^2$, and
has the unusual feature that no regularization of the second
functional derivative term in $H\/$ is required.

We study the contribution to
\begin{math}
  \langle F [G] | M | F [G] \rangle
\end{math}
from diagonal metrics
\begin{equation}
  \label{eq:gijx}
  G_{ij} (x) =
  \left(
    \begin{array}{cc}
      \lambda (r) & 0 \\
      0 & 1
    \end{array}
  \right),
\end{equation}
a restriction made just to simplify calculations.  We assume the
$C_\infty\/$ form
\begin{math}
  \lambda (r) = r^2 f (r^2)
\end{math},
with $f (0) \neq 0$ so that $G_{ij} (x)$ is degenerate at $r =
0$.  A simple calculation gives the non-vanishing Christoffel
symbols and scalar curvature
\begin{equation}
  \label{eq:gammaR}
  \renewcommand{\arraystretch}{2}
  \everymath{\displaystyle}
  \begin{array}{c}
    \Gamma^1_{11} = \half \frac x r (\ln \lambda)' \quad
    \Gamma^1_{12} = \half \frac y r (\ln \lambda)' \quad
    \Gamma^2_{11} = - \half \frac y r \lambda' \\
    R = -
    \left\{
      \frac 1r
      \left(
        1 - \frac{y^2}{r^2}
      \right) (\ln \lambda)' - \frac{y^2}{r^2}
      \left[
        (\ln \lambda)'' + \half
        \left(
          ( \ln \lambda )'
        \right)^2
      \right]
    \right\}.
  \end{array}
\end{equation}
The curvature behaves as $1/r^2$ at the origin, so the term
\begin{equation}
  \label{eq:intdsqre}
  \int d^2 x (\det G) R^2 F^* F
\end{equation}
gives a logarithmic divergent contribution to the energy which is
not canceled elsewhere.  (In this case the Christoffel symbols
are not singular enough to make the
\begin{math}
  \left|
    \nabla \frac{\partial F}{\partial G}
  \right|^2\/
\end{math}
terms diverge.)

Of course an integral over all metrics is required to compute
the expectation value of the energy $\langle F | H | F \rangle$.
It is possible that the infinity found above is irrelevant if the
``total functional measure'' of degenerate metric configurations
vanishes.  We will attempt to address this issue in Sec.~V, and
we now turn to a discussion of the analogous quantum mechanical
situation.

\section{Quantum Mechanical Central Force Problem for
\protect\boldmath$d = 2$} \label{sec:4}

For reasons that will become clear very quickly, the central
force problem in two space dimensions is the relevant quantum
mechanical analog for the problem of energy barriers in gauge
field theory.

We study
\begin{equation}
  \label{eq:study}
  \angles\psi H \psi =
  \int d^2\! x
  \left[
    \nabla \psi^* \cdot \nabla \psi + V (r) \psi^* \psi
  \right]
\end{equation}
for a central potential \( V(r)\/\) which is non-singular at \( r
= 0\).  It is useful to rewrite this in the form
\begin{equation}
  \label{eq:form}
  \angles \psi H \psi =
  \int d^2\! x
  \left[
    \half
    \left|
      ( \partial_x + i \partial_y ) \psi
    \right|^2
    + \half
    \left|
      (\partial_x - i \partial_y) \psi
    \right|^2 + V (r) \psi^* \psi
  \right]
\end{equation}
and introduce the wave function
\begin{equation}
  \label{eq:function}
  \psi = e^{i m \theta} f(r) =
  \left(
    \frac{x + i y}{r}
  \right)^m f (r).
\end{equation}
After calculating derivatives and doing the angular integral, one
obtains
\begin{equation}
  \label{eq:obtains}
  \angles \psi H \psi =
  2 \pi \int^\infty_0 r\, dr
  \left\{
    \half
    \left|
      f' - \frac m r f
    \right|^2 + \half
    \left|
      f' + \frac mr f
    \right|^2 + V (r) |f|^2
  \right\}.
\end{equation}
There are two possibly singular barrier terms, and the energy is
infinite unless both conditions
\begin{equation}
  \label{eq:cond}
  \lim_{r \to 0} r
  \left|
    f' \mp \frac mr f
  \right| = 0
\end{equation}
hold.  This gives only the very weak vanishing condition \( f (r)
\sim r^\epsilon\).

To obtain a stronger condition we assume that the radial wave
function has the product structure \( f (r) = f_c (r) R (r)\/\)
where \( f_c (r)\/\) satisfies the equation
\begin{equation}
  \label{eq:eq}
  f'_c - \frac mr f_c = 0
\end{equation}
with solution \( f_c (r) = r^m\).  For \( m > 0\/\) this vanishes
at the origin, so the \( f' + \frac mr f\/\) barrier condition is
satisfied in the limit \( r \to 0\).  If \( m < 0\/\) the roles
of the two conditions are reversed.  The net result is the
statement that
\begin{equation}
  \label{eq:that}
  f (r) = f^{|m|} R (r)
\end{equation}
with no constraints on the regular function \( R (r)\).  Of
course we made the extra assumption (\ref{eq:eq}) in order to
apply the barrier analysis to the first order form of $H$, but
the final result (\ref{eq:that}) agrees with the more rigorous
analysis of the second order Schr\"odinger equation.

One should note that for \( d = 3\), the radial measure is \( \int
r^2\, dr\/\) while the barrier singularity is again \( 1 / r^2\),
so it does not seem possible to apply barrier analysis to the
first order form of $H$.


\section{The Barrier Functional}
\label{sec:5}

In this section we develop the analogy between degenerate
configurations of the tensor \( G_{ij} (x)\/\) or the
electric field \( E^{a i} (x)\), and the singular point \( r
= 0\/\) in quantum mechanics.  We show that all physical wave
functionals have the representation
\begin{equation}
  \label{eq:PsiE2}
  \Psi [E] = e^{i \Omega [E]} F_c [G_{ij}] {\cal R} [G_{ij}],
\end{equation}
and define the centrifugal functional \( F_c [G_{ij}]\/\) which
``takes care of'' the singularities discussed in
Sec.~\ref{sec:singularity}, either by vanishing for degenerate
fields, and/or leaving a less singular functional Schr\"odinger
equation for the residual factor \( {\cal R} [G_{ij}]\).

The first step is to note that all states which satisfy the Gauss
law constraint carry the phase factor \( e^{i \Omega [E]}\/\)
which is the analogue of the angular factor \( e^{i m \theta}\)
or \( Y^m_\ell (\hat x)\/\) for non-zero angular momentum waves
in quantum mechanics.  These angular functions are singular at \(
r = 0\), specifically they are not continuous functions of the
Cartesian coordinates $x$, $y$, $z\/$ at the origin.  The phase
\( \Omega [E]\/\) has a similar behavior for degenerate fields
which is most easily seen using the following canonical
parametrization of the rectangular electric field matrix \(
E^{ai} (x)\), or, equivalently, its dual ${e^a}_j$.  If \( T^{a1}
(x)\/\) and \( T^{a 2} (x)\/\) are two orthogonal 3-vectors, \(
\mu_1 (x)\/\) and \( \mu_2 (x)\/\) non-negative real functions
with \( \mu_1 (x) < \mu_2 (x)\), and \( {\cal R}_{a i} = \cos \theta (x)
\delta_{\alpha i} - \sin \theta (x) \epsilon_{\alpha i}\/\) is a
$2 \times 2$ orthogonal matrix, then the frame \( {e^a}_j\), can
be expressed as
\begin{equation}
  \label{eq:eaj}
  {e^a}_j (x) =
  \sum^2_{\alpha = 1} T^{a \alpha} (x) \mu_\alpha (x)
  {\cal R}_{\alpha i} (x),
\end{equation}
a product of ``gauge, eigenvalue and spatial rotation'' parts.
This is essentially the dimensional reduction of the
parametrization of the square electric field matrix \( E^{a
i}\/\) used in the \( 3 + 1\/\) dimensional case in \cite{GJ1}.
If we substitute this parametrization into the representation
(\ref{eq:2.30a}) for the phase \( \Omega [E]\), one finds
\begin{equation}
  \label{eq:OmegaE}
  \Omega [E] =
  \int d^2x \epsilon^{i j} \sum^2_{\alpha = 1} e^a
  \left(
    \partial_i T^{a \alpha}
  \right) \mu_\alpha {\cal R}_{2 j}
\end{equation}
with \( e^a = \epsilon^{a b c} T^{b1} T^{b2}\).  At a point of
degeneracy, where \( \mu_1 (x_0) = 0\), the frame behaves as
\begin{equation}
  \label{eq:eajx0}
  {e^a}_j (x_0) \longrightarrow T^{a 2} (x_0) \mu_2 (x_0)
  {\cal R}_{\alpha i} (x_0)
\end{equation}
which is independent of the first row \( T^{a1} (x_0)\/\) of the
``gauge matrix,'' but the integrand of (\ref{eq:OmegaE}) still
depends on \( T^{a 1} (x_0)\).  It is this behavior which is
qualitatively similar to \( e^{i \theta}\/\) and \( Y^m_\ell
(\hat x)\).

The next question we ask is whether \( \angles F H F\/\) in
(\ref{eq:2.53}) is singular enough to permit a ``first order barrier
analysis.''  The clearest way we presently know to address this
question is to use a discretization of our Hamiltonian,
specifically a rectangular lattice with replacement of spatial
derivatives by discrete derivatives.  To justify this we recall
that one of the principal arguments for a reformulation of
non-abelian gauge theory in gauge invariant variables, is that
with such variables a cutoff has a gauge invariant meaning.  So
the crude lattice cutoff we use here should be satisfactory, and
we provisionally adopt the attitude that if there is an infinite
energy barrier problem in the discretized theory, then it is also
a significant issue in the continuum.

It is technically cleaner to study the singularities of \( H\/\)
for degenerate metrics, using the parameterization which follows
from (\ref{eq:eaj}), namely
\begin{equation}
  \label{eq:GijI}
  G_{ij} (I) = \sum_{\alpha = 1, 2} {\cal R}_{i \alpha} (I)
  \lambda_\alpha (I) {\cal R}_{j \alpha} (I)
\end{equation}
where \( I\/\) refers to the lattice site and \( \lambda_\alpha
(I) = (\mu_\alpha (I))^2\), \( 0 \leq \lambda_1 (I) \leq
\lambda_2 (I)\).  At each lattice site one has the chain rule and
measure
\begin{equation}
  \label{eq:frac}
  \everymath{\displaystyle}
  \renewcommand{\arraystretch}{2}
  \begin{array}{rcl}
    \frac{\delta}{\delta G_{11}} & = &
    \cos^2 \theta \frac{\delta}{\delta \lambda_1} + \sin^2 \theta
    \frac{\delta}{\delta \lambda_2} + \frac{\sin \theta \cos
      \theta}{\lambda_2 - \lambda_1} \frac{\delta}{\delta \theta}
    \\
    \frac{\delta}{\delta G_{22}} & = &
    \sin^2 \theta \frac{\delta}{\delta \lambda_1} + \cos^2 \theta
    \frac{\delta}{\delta \lambda_2} - \frac{\sin \theta \cos
      \theta}{\lambda_2 - \lambda_1} \frac{\delta}{\delta \theta}
    \\
    \frac{\delta}{\delta G_{12}} & = &
    \sin \theta \cos \theta
    \left(
      \frac{\delta}{\delta \lambda_1} - \frac{\delta}{\delta
        \lambda_2}
    \right) + \frac{\cos^2 \theta - \sin^2 \theta}{4 (\lambda_2 -
      \lambda_1)} \frac{\delta}{\delta \theta}
    \\
    \prod_{i \leq j} d G_{ij} & = &
    \left(
      \lambda_2 - \lambda_1
    \right) d \lambda_1\, d \lambda_2\, d \theta.
  \end{array}
\end{equation}
The chain rule can simply be substituted in the lattice
Hamiltonian.  One can show that the (discretized) Christoffel
symbols and scalar curvature behave like \( 1 / \lambda_1 (I)\/\)
(to within \( \log \lambda_1 (I)\/\) terms) at a point of
degeneracy of the configuration \( \left\{ G_{ij} (I)
\right\}\/\) of the discretized metric.  Putting things together
we find that both the \( | \delta F / \delta G |^2\/\) and
\begin{math}
  \left|
    R F + \delta^2 F / \delta G^2
  \right|^2
\end{math}
terms of \( H\/\) contain the effective barrier singularity \( d
\lambda_1 (I) / \lambda_1 (I)\/\) at each site.  This singularity has
the same strength as in the \( d = 2\) quantum mechanics problem,
so the energy is infinite unless wave functions are specially
constrained as \( \lambda_1 (I) \to 0\) at any site.  Specially
constrained does not necessarily mean that wave functions vanish,
but we prefer to discuss the situation further in the continuum
language.

Before doing this we would like to discuss the apparent
singularity when \( \lambda_1 (I) = \lambda_2 (I)\/\) at any lattice
site.  The net strength of the combined measure and singular
terms from the chain rule is \( 1 / \left| \lambda_2 (I) -
\lambda_1 (I) \right|\), so there is again a potential infinite
energy problem.  However we believe that this singularity is an
artefact of the choice of variables which is resolved with no
physical effect.  Specifically the singularity, which originates
in the chain rule (\ref{eq:frac}), is immediately cancelled if
the \( \delta / \delta \theta\/\) derivative acts on functionals
\( F [G_{ij}]\), where the \(\theta\) dependence appears only via
the metric components in (\ref{eq:GijI}).

We cannot yet formulate a precise criterion to distinguish
between singularities of possible physical significance and those
which are just mathematical artifacts.  We believe that the
physical singularities are those of the phase \( \Omega [E]\/\)
which can be expressed as gauge invariant
statements about the electric field configuration.  An optimal
choice of gauge invariant variables is one in which no further
singularities appear in the chain rule.  This is true for the
metric variables \( G_{ij} (x)\).

We have reached the conclusion that the singularities of \( H\/\)
for degenerate metrics are significant enough to place possibly
interesting constraints on wave functionals.  Since the situation
is similar to \( d = 2\/\) quantum mechanics, we shall try to
apply the barrier analysis of Sec.~\ref{sec:4} to our Hamiltonian
in the ``effective first order form'' by which it is given in
(\ref{eq:2.53}).  We must require that physical wave functionals
\( F [G]\/\) satisfy
\begin{eqnarray}
  \label{eq:nabla1}
  \nabla_j \frac{\partial F}{\partial G_{ij} (x)} & = &
  \mbox{smooth} \\
  \label{eq:epsilon}
  \epsilon_{ik} \epsilon_{j \ell}
  \frac{\partial^2 F}{\partial G_{ij} (x)\, \partial G_{k \ell}
    (x)} + \frac 14 R (x) F & = & \mbox{smooth}
\end{eqnarray}
where ``smooth'' means less singular than \( \Gamma^i_{jk}
(x)\) or \( R (x)\) at points where \( G_{ij} (x)\) is
degenerate.

There is a simple qualitative interpretation of
(\ref{eq:nabla1}), since if ``= smooth'' is replaced by ``= 0,''
we simply have the condition that \( F [G]\/\) is a
diffeomorphism invariant functional of $G_{ij}$.  Of course the
full wave functional of the gauge theory cannot be diffeomorphism
invariant since \( H\/\) does not have this property.  But
qualitatively one has the picture of a functional for which the
violation of diffeomorphism invariance is ``soft'' near
degenerate metrics.  The functional
\begin{equation}
  \label{eq:FandG}
  F [G] = \exp - \int d^2 x\, \sqrt G
  \left[
    1 + \delta^{ij} G_{ij}
  \right]
\end{equation}
is one example.  There is also a definite qualitative
interpretation of (\ref{eq:epsilon}), which we discuss below, but
we note that we have not been able to find any explicit
functional which satisfies (\ref{eq:epsilon}).  What we have
discussed so far is a ``conservative'' approach to the barrier
singularities, and this approach must be called a failure, since
it has produced only a weak and vague picture.

Therefore we shall be bolder and postulate the product structure
(\ref{eq:PsiE2}), with \( F_c [G]\/\) defined as the solution of
the equations
\begin{eqnarray}
  \label{eq:nablai}
  \nabla_j \frac{\delta F_c}{\delta G_{ij} (x)} & = & 0 \\
  \label{eq:nab}
  \epsilon_{ik} \epsilon_{j \ell}
  \frac{\delta^2 F_c}{\delta G_{ij} (x)\, \delta G_{k\ell} (x)} +
  \frac 14 R (x) F_c & = & 0
\end{eqnarray}
One question to ask is whether the three conditions are mutually
compatible, since it was the incompatibility of the two
conditions (\ref{eq:cond}) which led to the condition that the
radial wave function vanishes at the origin.  It turns that
(\ref{eq:epsilon}) and (\ref{eq:FandG}) are compatible, since it
is precisely these equations that emerge from a rather different
physical context.  They are the diffeomorphism and
Wheeler-de-Witt constraints of the metric formulation of \( 2 +
1\/\) dimensional general relativity \cite{SC} (after
continuation of the time coordinate to Euclidean signature).  The
quantum theory of \( 2 + 1\/\) dimensional gravity has been
widely studied, but the usual procedure is to reduce to the
finite number of degrees of freedom of a topologically
non-trivial compact spatial manifold.  Our \( F_c [G_{ij}]\/\) is
the unreduced wave functional which is expected to be the unique
physical state for the topologically trivial situation of a
non-compact spatial 2-surface, and we are not aware of any known
explicit solution.

More progress can be made if we consider the equivalent
Chern-Simons \cite{AT1} (or topological \( b/F\/\) theory \cite{H2}
) with
action
\begin{equation}
  \label{eq:S}
  \everymath{\displaystyle}
  \renewcommand{\arraystretch}{2}
  \begin{array}{rcl}
    S & = & \half \int d^3 x \epsilon^{\lambda \mu \nu}
    e^a_\lambda F^a_{\mu \nu} \\
    & = & \int d^3 x \epsilon^{\lambda \mu \nu} e^a_\lambda
    \left(
      \partial_\mu A^a_\nu - \partial_\nu A^a_\mu + e^{a b c}
      A^b_\mu A^c_\nu
    \right).
  \end{array}
\end{equation}
One observes that \( A^a_i (x)\/\) and \( E^{a i} (x) \equiv
\epsilon^{ij} e^a_j (x)\/\) are canonically conjugate variables
which satisfy (\ref{eq:2.2}), and that the physical state
functional \( \Psi_c [E]\/\) in (the unconventional) $E\/$-field
representation, satisfies the constraint equations
\begin{eqnarray}
  \label{eq:Di}
  D_i E^{a i} (x) \Psi_c [E] & = & 0 \\
  \label{eq:Ba}
  B^a (x) \Psi_c [E] & = & \half \epsilon^{i j} F^a_{ij} (x)
  \Psi_c [E] = 0.
\end{eqnarray}
The first of these is just the Gauss law constraint
(\ref{eq:2.3}) of the non-abelian gauge theory, while
(\ref{eq:Ba}) is entirely equivalent to (\ref{eq:nablai}) and
(\ref{eq:nab}) together.  To see this, one need only note from
the form of the Hamiltonian that the constraints
(\ref{eq:nablai}) and (\ref{eq:nab}) can be interpreted before
the unitary transformation (\ref{eq:2.11}) as the simple
statement that the magnetic field \( B^a (x)\/\) annihilate the
state \( \Psi_c [E] = e^{i \Omega [E]} F_c [G]\).  This is just
the state which corresponds to the singular object
\begin{equation}
  \label{eq:Psic2}
  \Psi_c [A] = \prod_{a, x} \delta
  \left(
    B^a (x)
  \right)
\end{equation}
in connection representation.

It is a fairly straightforward exercise \cite{MS1} to show
that
\begin{equation}
  \label{eq:PsiE3}
  \everymath{\displaystyle}
  \renewcommand{\arraystretch}{2}
  \begin{array}{rcl}
    \Psi_c [E] & = & \int [dA] \exp
    \left(
      -i \int d^2 x\, E^a (x) A^a_i (x)
    \right) \Psi_c [A] \\
    & = & \int [dU] \exp \int d^2 x \Tr
    \left(
      E^i U^{-1} \partial_i U
    \right)
  \end{array}
\end{equation}
where \( U (x)\/\) is a \( 2 \times 2\/\,\) SU(2) matrix and \(
E^i = {\cal T}^a E^{a i}\), with the Pauli matrices \( {\cal
T}^a\).  One can also show by direct functional differentiation
that (\ref{eq:PsiE3}) satisfies (\ref{eq:Ba}), essentially
because \( U^{-1} \partial_i U\/\) is a ``pure gauge.''

The verification of (\ref{eq:Di}) is less direct but useful for
the further development.  We note that the phase \( \Omega
[E]\/\) can be written in matrix form as
\begin{eqnarray}
  \label{eq:array}
  \Omega [E] & = &
  - \half \int d^2 x\, \Tr
  \left(
    E^i \omega_i
  \right)
\end{eqnarray}
with \( \omega_i = {\cal T}^a \omega^a_i\), a form which can be
obtained by substitution of (\ref{eq : cov}) in (\ref{eq:2.30a}).
This means that \( \Psi_c [E]\/\) can be rewritten as
\begin{equation}
  \label{eq:EcPsi}
  \Psi_c [E] =
  \exp i \Omega [E] \int dU(x)\, \exp
  \int d^2 x \Tr
  \left[
    E^i U^{-1}
    \left(
      \partial_i U + \frac i 2 U \omega_i
    \right)
  \right]
\end{equation}
and the second factor can easily be shown to be invariant under
the gauge transformations
\begin{equation}
  \label{eq:Esquared}
  E^i \longrightarrow V^{-1} E^i V \qquad U \to U V.
\end{equation}
Thus \( \Psi_c [E]\/\) is the product of the phase factor \( \exp
i \Omega [E]\/\) times an explicitly gauge invariant functional
of \( E^i\).  According to the original argument of
Sec.~\ref{sec:2}, this means that \( \Psi_c [E]\/\) satisfies the
Gauss constraint (\ref{eq:Di}).

The same argument also tells us that the second factor in
(\ref{eq:EcPsi}) gives a functional integral representation of
the centrifugal functional \( F_c [G]\/\) which satisfies
(\ref{eq:nablai}) and (\ref{eq:nab}).  Another form of this
can be obtained by writing
\begin{equation}
  \label{eq:Ux}
  U (x) =
  \left(\!\!
    \begin{array}{rr}
      u^*_1 (x) & u^*_2 (x) \\
      -u_2 (x)  & u_1 (x)
    \end{array}\!\!
  \right) \qquad
  u^*_\alpha u_\alpha = 1.
\end{equation}
Inserting this in (\ref{eq:EcPsi}) we find after some
manipulation the representation
\begin{equation}
  \label{eq:mani}
  F_c [G] = \int
  \left[
    du_\alpha (x)\, du^*_\beta (x)\, \delta
    \left(
      u^*_\gamma (x) u_\gamma (x) - 1
    \right)
  \right] \exp
  \left[
    -2 \int d^2x\, u^* E^i D_i u
  \right]
\end{equation}
where
\begin{equation}
  \label{eq:diu}
  D_i u =
  \left(
    \partial_i - \frac i 2 {\cal T}^a \omega^a_i
  \right)
  \left(
    {u_1 \atop u_2}
  \right).
\end{equation}
This involves a Dirac-like operator and a non-linear measure on
the ``spinor'' fields which reflects the original SU(2)
constraints on the matrix \( U(x)\).  The argument of the
exponential in (\ref{eq:mani}) is imaginary, since the
differential operator is anti-Hermitean,
\begin{equation}
  \label{eq:intdsq}
  \int d^2x\,
  \left(
    u^* E^i D_i u
  \right)^{\!*} = - \int d^2 x
  \left(
    u^* E^i D_i u
  \right)
\end{equation}
(where (\ref{eq:2.26}) has been used).

It has not been shown explicitly that (\ref{eq:mani}) is a
functional of \( G_{ij} (x)\), but it follows from the arguments
of Sec.~\ref{sec:2} that a gauge invariant functional of \( E^{a
i} (x)\) depends only on \( G_{ij} (x)\).  Some support for these
arguments and the additional fact that \( F_c [G]\) is real comes
if we temporarily denote (\ref{eq:mani}) by \( F_c [E]\).  One
then has
\begin{equation}
  \label{eq:Fcstar}
  F_c [E]^* = F_c [-E] = F_c [E].
\end{equation}
The first equality requires only (\ref{eq:intdsq}) and the fact
that \( \omega_i\/\) is even under \( E^i \to - E^i\), and the
second follows from the observation that field configurations \(
E^i (x)\/\) and \( -E^i (x)\/\) are related by the gauge
transformation which describes a rotation by \( 180^\circ\/\)
about the axis \( e^a \sim \epsilon^{a b c} E^{b 1} E^{c 2}\).
It is also worth observing that any pair of 3-vectors \( E^{a 1}
(x)\/\) and \( E^{a 2} (x)\/\) can be gauge rotated to the 1--2
plane; \( E^{31} (x) = E^{32} (x) = 0\/\) and \( e^a (x) = (0, 0,
\pm 1)\).  The frame \( e^a_i = -\epsilon_{ij} E^{a j}\/\) can
then be viewed as a standard zweibein for the metric \(
G_{ij}\), and the connection \( \omega^a_i\/\) is related to the
standard spin connection by \( \omega^a_i = - \half
\epsilon^{abc} \omega^{bc}_i\).  In this (partially fixed) gauge
the differential operator in (\ref{eq:mani}) is just the
standard Dirac operator for the 2-manifold with zweibein \( e^a_i
(x)\/\) and Riemannian spin connection.

We have implicitly assumed that the functional integral
representations for \( \Psi_c [E]\/\) and \( F_c [G]\/\) are well
defined despite the fact that they involve oscillating
integrands, and we hope this is true because \( \Psi_c [E]\/\) is
the wave functional of a topological field theory.  It is also to
be hoped that the knowledge of two-dimensional and topological
field theories that has developed during the last decade of work
in mathematical physics will lead to some progress toward the
evaluation of these path integrals.  One approach is to consider
a semi-classical approximation constructed using classical
solutions of either the metric or topological formulations of \(
2 + 1\/\) dimensional gravity.  Although the approximation may
not satisfy (\ref{eq:nablai}) or (\ref{eq:nab}) exactly, the
smoothness conditions (\ref{eq:nabla1}) and (\ref{eq:epsilon})
may be satisfied, and that could be sufficient for the purposes
of gauge field theory.

The next step in the exploration of the consequences of the
energy barrier is to substitute the product $F [G] = F_c [G]
{\cal R} [G]\/$ in the Hamiltonian (\ref{eq:2.53}) in order to
study the effective Hamiltonian governing the residual factor
${\cal R} [G]\/$ in (\ref{eq:PsiE2}).  Using (\ref{eq:nablai})
which expresses the diffeomorphic invariance of $F_c [G]$, it is
easy to see that
\begin{equation}
  \label{eq:5.24}
  \nabla_j \frac{\delta}{\delta G_{j \ell}}
    \left(
      F_c [G] \, {\cal R} [G]
    \right) =
    F_c [G] \nabla_j \frac{\delta {\cal R}}{\delta G_{j \ell}}.
\end{equation}
It is also straightforward to apply the second functional
derivative to the product, and then use (\ref{eq:nab}) to obtain
\begin{eqnarray}
  2 \epsilon_{i k} \epsilon_{j \ell} \frac{\delta^2 F}{\delta
    G_{ij} \delta G_{k \ell}} + \half {\cal R} F & = &
  2 F_c [G] \epsilon_{ik} \epsilon_{j \ell}
  \left[
    \frac{\delta^2 {\cal R}}{\delta G_{ij} \delta G_{k \ell}} +
    2 \rho_{ij} \frac{\delta {\cal R}}{\delta G_{k \ell}}
  \right] \nonumber \\
  \label{eq:5.25}
  \rho_{ij} (x) & \equiv &
  \frac{\delta}{\delta G_{ij} (x)} \ln F_c [G].
\end{eqnarray}
We then substitute these results in (\ref{eq:2.53}) obtain the new
form of the Hamiltonian:
\begin{equation}
  \label{eq:5.26}
  \everymath{\displaystyle}
  \begin{array}{rcl}
    \angles F H F & = & \half \int d^2 x \int [d G_{ij}]\,
    F^2_c [G] \\
    && \quad \strut
    \left\{
      \vphantom{
        \left|
          \epsilon_{i h} \epsilon_{j 0}
          \left(
            \frac{\delta^2 {\cal R}}{\delta G_{ij} \delta G_{k \ell}} +
            2 \rho_{ij} \frac{\delta {\cal R}}{\delta G_{k \ell}}
          \right)
        \right|^2}
      g^2 \delta^{ij} G_{ij} {\cal R}^* {\cal R}
    \right. + \frac{4 G_{k \ell}}{g^2} \nabla_i
    \frac{\delta {\cal R}^*}{\delta G_{i k}} \nabla_j
    \frac{\delta {\cal R}}{\delta G_{j \ell}} \\
    && \qquad
    \left.
      + \frac{4 \det G}{g^2}
      \left|
        \epsilon_{i k} \epsilon_{j l}
      \left(
        \frac{\delta^2 {\cal R}}{\delta G_{ij} \delta G_{k \ell}} +
        2 \rho_{ij} \frac{\delta {\cal R}}{\delta G_{k \ell}}
      \right)
      \right|^2
    \right\}
  \end{array}
\end{equation}

Let us discuss this result, noting first that the expression
within the brackets \{ \ \} is less problematic for degenerate
fields than the analogous term in (\ref{eq:2.53}) because the
singular quantity $R F$ has been removed.  Indeed if the
prefactor $F_c [G]^2\/$ vanishes, then one expects no special
constraints on ${\cal R} [G]$.  However we must also entertain
the possibility that $F_c [G]\/$ does not vanish, for degenerate
fields.  If its logarithmic derivative $\rho_{ij} (x)\/$ is also
regular, then the only constraint on ${\cal R} [G]\/$ comes from
the diffeomorphism term in (\ref{eq:5.26}).  One can avoid an
infinite contribution to the energy if, as discussed earlier in
this section, ${\cal R} [G]\/$ is a functional with a ``soft''
violation of diffeomorphism invariance near degenerate metrics.
One may also have the situation that $F_c [G]\/$ is non-vanishing
and $\rho_{ij}\/$ is singular.  In this case, it is difficult to
be precise, but we expect that the constraint on ${\cal R} [G]\/$
from the $\epsilon_{ij} \epsilon_{j \ell}\/$ term is less severe
than for $F\/$ itself because there is no longer a singular
purely multiplicate term like $R F$.

Although our investigation has ended in an indefinite way, it is
worth summarizing the line of thinking presented in this section.
We started by considering the singularities of a formally correct
Hamiltonian for a non-abelian gauge theory.  Working by analogy
with quantum mechanics, we were led to postulate the product
structure (\ref{eq:PsiE2}) for physical state functionals, and we
found that the barrier functionals \( \Psi_c [G]\/\) or \( F_c
[G]\/\) have a direct interpretation in a simple topological
field theory.  It also appears that the factorization of $F_c
[G]\/$ leaves a less singular effective Hamiltonian for ${\cal R}
[G]\/$ whether or not $F_c [G]\/$ vanishes for degenerate fields.
The product structure (\ref{eq:PsiE2}) is entirely correct, but
whether it is useful or not requires further information about
the barrier functional $F_c [G]$.

\section{Acknowlegement} \label{sec:6}

The authors thank L.~Alvarez-Gaum\'e (CERN), R.~Brooks,
S.~Carroll, K.~Johnson, G.~Lifshytz, S.~Mathur, and B.~Zwiebach
(MIT) for several useful discussions.  Special thanks go to
M.~Ortiz (Tufts) for the important observation that
(\ref{eq:nab}) is just the Wheeler-de-Witt equation of \(2 + 1\)
dimensional gravity!  This research began when both authors
enjoyed the support of TH-Division CERN.

\end{document}